\documentclass[prl,twocolumn,showpacs,preprintnumbers,amsmath,amssymb]{revtex4-1}

\usepackage{graphicx}
\usepackage{ifthen}
\usepackage{ifpdf}
\usepackage{color}
\definecolor{red}{rgb}{1.,0.,0.}

\usepackage{latexsym}
\usepackage{amssymb}
\usepackage{bm}
\newcommand{\half}{\mbox{\small $\frac{1}{2}$}}

\newcommand{\eexp}{\mbox{e}^}

\newcommand{\beq}[1]{\begin{eqnarray}\ifthenelse{#1=-1}{\nonumber}
{\ifthenelse{#1=0}{}{\label{e#1}}}}
\newcommand{\eeq}{\end{eqnarray}}
\newcommand{\be}[1]{\begin{eqnarray}}
\newcommand{\ee}{\end{eqnarray}}

\newcommand{\hide}[1]{}

\begin{document}

\title{Phase transitions for a collective coordinate coupled to Luttinger liquids}

\author{B. Horovitz}
\affiliation{ Department of Physics, Ben Gurion University,
Beer Sheva 84105 Israel}
\author{T. Giamarchi}
\affiliation{DPMC-MaNEP, University of Geneva, 24 Quai Ernest Ansermet, 1211 Geneva 4, Switzerland}
\author{P. Le Doussal}
\affiliation{CNRS-Laboratoire de
Physique Th{\'e}orique de l'Ecole Normale Sup{\'e}rieure, 24 rue
Lhomond,75231 Cedex 05, Paris France.}
\date{\today}

\begin{abstract}
We study various realizations of collective coordinates, e.g. the position of a particle, the charge of a Coulomb box or the phase of a Bose or a superconducting condensate, coupled to Luttinger liquids (LL) with $N$ flavors. We find that for a Luttinger parameter $\half<K<1$ there is a phase transition from a delocalized phase into a phase with a periodic potential at strong coupling. In the delocalized phase the dynamics is dominated by an effective mass, i.e. diffusive in imaginary time, while on the transition line it becomes dissipative. At $K=\half$ there is an additional transition into a localized phase with no diffusion at zero temperature.
\end{abstract}

\maketitle

Diffusion and propagation of massive particles surrounded by a bath
is one of very challenging problem of condensed matter.
Historically it started with the celebrated Brownian motion \cite{risken_fokker_planck}
in which the interactions between a classical particle and the
microscopic motion of the classical bath, lead to a diffusion,
connected by the Einstein relation to a finite friction.

This problem gets incredibly more complicated when the bath becomes
quantum. Indeed the excitations of the bath can lead, by Anderson orthogonality
effects to a modification of the motion of the quantum particle or the collective coordinate coupled
to the bath \cite{leggett_two_state}.
One of the realization of such problem is the polaron problem \cite{feynman_statmech}
where the interaction with the vibrations of the lattice leads to
an increase of the mass of the particle and even potentially to self trapping.
This type of problem has recently benefitted from the recent progress in cold atomic systems
\cite{bloch_cold_atoms_optical_lattices_review}. Indeed in such systems impurities in quantum baths can be realized
in a variety of manners ranging from fermi- or bose- mixtures to ions in condensates, and at various dimensionalities
\cite{schirotzek_imbalanced_polaron,nascimbene_imbalanced_polaron,gadway_mixtures_impurities,%
spethmann_impurity_cs,zipkes_ion_condensate,schmid_dynamics_ion,chikkatur_impurity_velocity,palzer_impurity_1d,%
catani_impurity_oscillations,fukuhara_heinseberg_cold}.

A situation of special interest is provided by a one-dimensional bath for which the bath-bath correlations
can become highly non-universal, i.e. they acquire an interaction dependent powerlaw correlations characteristics
of a Luttinger liquid (LL) \cite{giamarchi_book_1d}. In that case special effects can potentially occur,
as clear from the static impurity case \cite{kane_qwires_tunnel} and mobile ones coupled to single baths \cite{castella_mobile_impurity,castroneto_mobile_impurity}. In particular
it was shown recently \cite{zvonarev_ferro_cold} that this led to a new universality class for the
motion of the impurity, for which in particular subdiffusion can occur. This very rich situation was explored
further. On the theory side both diffusive \cite{kamenev_exponents_impurity,zvonarev_gaudinyang,bonart_quantum_brownian}, kicked \cite{mathy_moving_impurity,massel_kicked_impurity} and driven impurities \cite{gangardt_mobile_impurity,lamacraft_mobile_impurity,schecter_mobile_impurity} were considered.
On the experimental side driven impurities \cite{palzer_impurity_1d}, mixtures of $^{87}$Rb and $^{41}$K \cite{catani_impurity_oscillations}
and $^{87}$Rb experiments with local addressability \cite{fukuhara_heinseberg_cold} were successful implementation of the one dimensional problem.

\begin{figure}[b]
\includegraphics[scale=0.4]{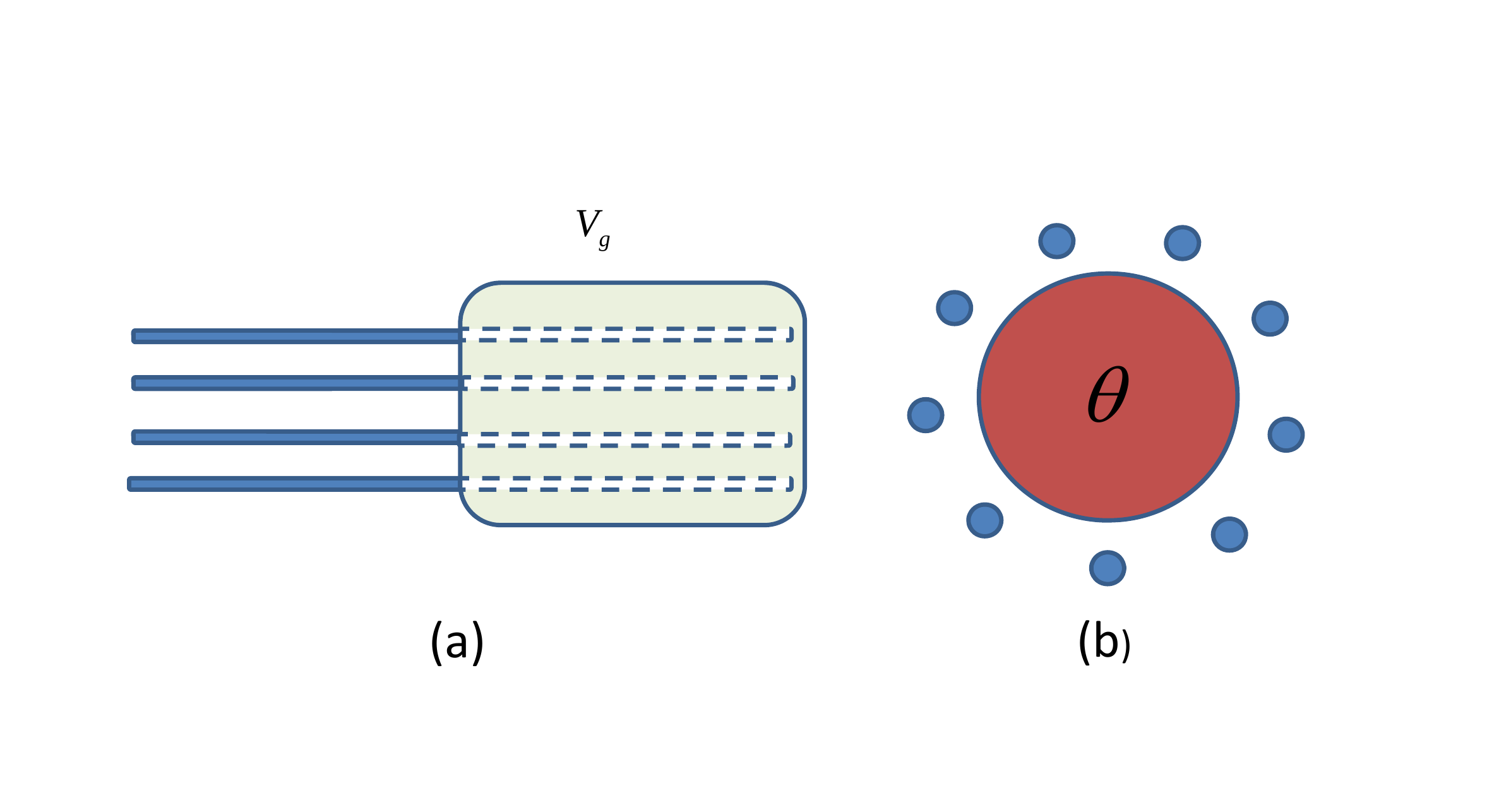}
\caption{\label{fig:experiments} Illustrations of collective coordinates coupled to LLs: (a) The environment of a few LLs enter a finite Coulomb box that is under a gate voltage $V_g$ and where the total charge interacts with an effective capacitance. (b) A BEC condensate with phase $\theta$ is Josephson coupled to bosonic LLs, shown by their cross section as they cross the figure plane. In a condensed matter context the figure could also represent a superconducitng grain coupled to one-dimensional superconducting wires, which will act as LLs.}
\end{figure}

\begin{figure}
\includegraphics[scale=0.8]{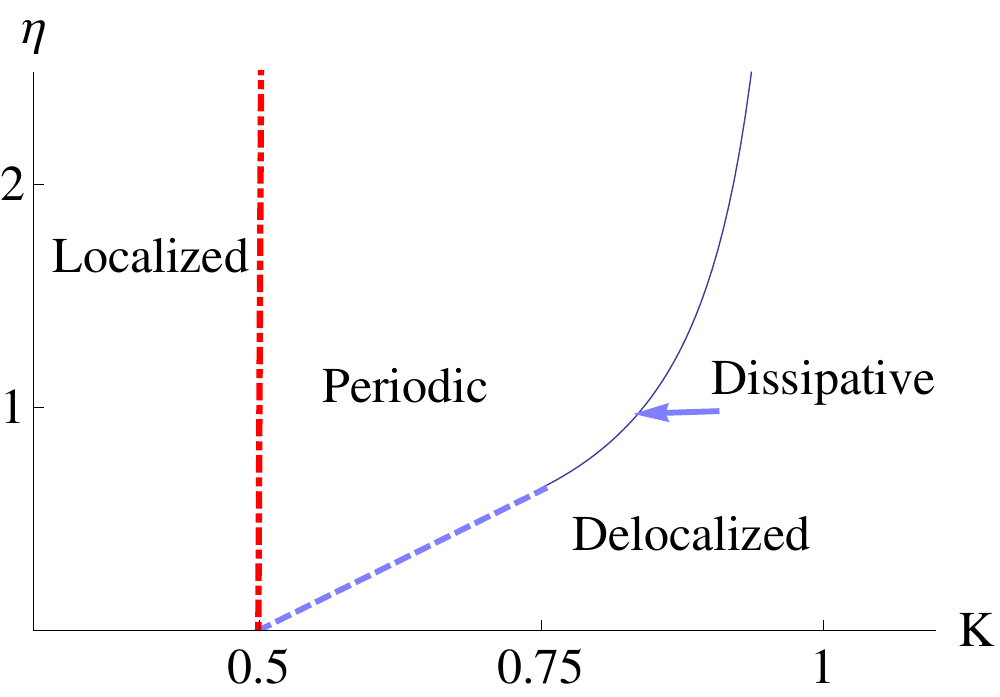}
\caption{\label{fig:phasediag} Phase diagram for an impurity in a bath of LLs as a function of the LL parameter $K$ and the interaction parameter between
the impurity and the bath $\eta$. Four regimes can occur (see text) in which the impurity is delocalized, just dissipative, periodically localized, or localized. Dashed lines indicates boundaries out of the control of the perturbative RG. The corresponding correlation functions are given in Table~\ref{tab:corr}.}
\end{figure}

In this paper we study the physics of a collective coordinate coupled to $N$ Luttinger baths ($N\gg 1$), e.g. a particle position, charge of a Coulomb box, a phase of a Bose-Einstein condensate (BEC), or a phase of a superconducting grain, see Fig.~\ref{fig:experiments}.
These potential experimental realizations are further examined before the conclusions.
We solve this system allowing for both LL density fluctuations and LL interaction and derive
 a novel localization-delocalization transition, as summarized in Fig.~\ref{fig:phasediag}. The localized and delocalized phases are separated by a line on which the motion is simply diffusive. We note that the collective coordinate represents a small system with correlations decaying in time. The periodic and localized phases are therefore of particular significance since the collective coordinate acquires long range order due to its interaction with the LLs.


For concreteness, the following presentation uses the particle coordinate language.
We consider a particle of mass $\tilde M$ coupled to a LL with a contact interaction $\tilde{H}_{\rm int} = g \rho(\tilde X)$ where $\tilde X$ is the operator measuring
the impurity position and $\rho(x)$ the density is the LL. We study this model in the large $N$ limit, so the impurity becomes coupled to $N$ independent LL and the interaction becomes $H_{\rm int} = g \sum_{i=1}^N \rho_i(\tilde X)$. The action of the system can be computed by a cumulant expansion in powers of $g$ and only the second order cumulant remains when $g^2N=O(1)$. Indeed the fourth order cumulant is of order $g^4N\sim 1/N$ and can be neglected. Using the expression of the density in a LL \cite{giamarchi_book_1d}
$\rho(x,\tau)=\rho_0-\frac{1}{\pi}\partial_x\phi(x,\tau)+\rho_0[\eexp{2i(\pi\rho_0x-\phi(x,\tau))}+h.c.]$ where $\phi(x,\tau)$ is the bosonic phase, and performing the Gaussian integration over the LL Hamiltonian, the action becomes
\begin{equation} \label{e01}
 S_{\rm eff}=\frac{M}2 \int_{\tau}(\dot X)^2_{\tau} \\ -\frac{\eta\Lambda^2}{2\pi}\int_{\tau}\int_{\tau'}\frac{\cos(X_{\tau}-X_{\tau'})}{(\Lambda(\tau-\tau'))^{2K}}
\end{equation}
where we have used the dimensionless variables $X = 2\pi \rho_0 \tilde X$ and $M = \tilde M /(2\pi\rho_0)^2$,
$\eta=2\pi g^2\rho_0^2 N/\Lambda^2$, $\tau$ is the imaginary time, $u$ the velocity of excitation in the LL, and $K$ the Luttinger parameter that controls the powerlaw decay of the correlation functions. A frequency cutoff $\Lambda=u/\alpha$ is used to have a dimensionless coupling $\eta$ where $\alpha\approx 1/\rho_0$ is the natural momentum cutoff of the LL. In the above expression only the oscillating (backscattering) term in the density has been retained. Indeed the $\partial_x\phi(x,\tau)$
interaction can be integrated, leading at long times to $(X_\tau-X_{\tau'})^2/(\tau-\tau')^4$, i.e. an $\omega^3$ term in frequency which can be neglected relative
to the bare kinetic energy term of the impurity $M\omega^2$. We have used that for a LL one obtains \cite{giamarchi_book_1d} $\langle \eexp{i\phi(X_\tau,\tau)-i\phi(X_{\tau'},\tau')}\rangle_{LL}\sim [(X(\tau)-X(\tau'))^2 + u^2 (\tau-\tau')^2]^{-K}$. We have also made the additional assumption, which will
be verified in what follows that the impurity is less than ballistic and thus that $\langle (X_\tau-X_{\tau'})^2\rangle\ll u^2(\tau-\tau')^2$.

To solve for the thermodynamics of (\ref{e01}) we consider first a renormalization group (RG) process \cite{kosterlitz_ising_longrange} valid for large $\eta$, which was also applied to the $K=1$ case \cite{guinea_flux_dissipative}. The action (\ref{e01}) is approximated by its short time form where it becomes Gaussian
\begin{equation} \label{e02}
 S_0 = \frac12\int_{\omega}[ M\omega^2+\eta C_K\Lambda^{2-2K}|\omega|^{2K-1}]|X(\omega)|^2
\end{equation}
where $\int_{\tau}\frac{1-\cos(\omega\tau)}{\tau^{2K}}=-2\Gamma(1-2K)\sin (K\pi) |\omega|^{2K-1}\equiv \pi C_K |\omega|^{2K-1}$ so that $C_K =1-0.85(K-1)+O(K-1)^2$. The cutoff $\Lambda$ is replaced by $\Lambda'$ and the interaction is averaged with $S_0$ in the small frequency interval $\Lambda'<\omega<\Lambda$ leading to $d\Lambda/\pi C_K\eta\Lambda$. The action has then a renormalized coefficient $\eta^R(\Lambda')^{2-2K}$ where
\begin{equation}\label{e03}
 \eta^R = \eta\left\{1+[(2-2K)-\frac{1}{\pi\eta}]\ln\frac{\Lambda}{\Lambda'}\right\}
\end{equation}
with $C_K\rightarrow 1$ to 1st order in either $1/\eta$ or $1-K$.
Hence if $K\geq 1$ $\eta^R$ flows to small values, while if $K<1$ there is an unstable fixed point at $\eta_c=
\frac{1}{2\pi(1-K)}$. $\eta>\eta_c$ flows to large
values, while $\eta<\eta_c$ flows to smaller values of $\eta$. One can integrate (\ref{e03}) when $\eta<\eta_c$ down to $\eta^R\approx 1$ below which the RG is not controlled. The new cutoff is interpreted as an effective mass \cite{guinea_flux_dissipative,hofstetter_blockade} $M^*$
\begin{equation} \label{e04}
 \frac{1}{M^*}\approx \Lambda [1-\pi\eta(2-2K)]^{\frac{1}{2-2K}}
\end{equation}
which for $\pi\eta(2-2K)\ll 1$ but $\pi\eta\gg 1$, i.e. far from the transition point, represents an exponentially large mass $M^*\sim \eexp{\pi\eta}$ as for the $K=1$ case \cite{guinea_flux_dissipative,hofstetter_blockade}.

To supplement this scenario, and study the properties of the three resulting phases, we follow a variational
scheme \cite{horovitz_aharonovbohm_dissipative} where we find the best quadratic action
approximating the original action (\ref{e01}). The corresponding Green's function $1/f(\omega)$ is a solution of the self-consistent equation
\begin{equation}\label{e05}
 f(\omega) = M\omega^2+\frac{2}{\pi}\eta\Lambda^{2-2K}\int_0^{\infty}d\tau\frac{1-\cos\omega\tau}{\tau^{2K}}
    \,\eexp{-\int_0^{\Lambda}\frac{1-\cos\omega'\tau}{\pi f(\omega')}}
\end{equation}
We note first that at $\omega\approx \Lambda$ the solution is $f(\omega)-M\omega^2\sim |\omega|^{2K-1}$. In the following we focus on $\omega\ll \Lambda$ and on $K<1$.
As a first option we consider $f(\omega) = \eta^*C_K \omega^{2K-1}/ \Lambda^{2K-2}$.  The integral in the exponent converges as $\tau\rightarrow \infty$, so it is $\int_0^{\Lambda} d\omega'/(\pi f(\omega')) = [\pi\eta^*C_K(2-2K)]^{-1}$, hence (\ref{e05}) reduces to
\begin{equation} \label{e06}
\eta^*=\eta \eexp{-[\pi\eta^* C_K(2-2K)]^{-1}}
\end{equation}
This equation has solutions only if $\eta$ is sufficiently large, i.e. $\pi C_K\eta(2-2K)>e$.
A second possible solution is $f(\omega)=\eta^*|\omega|$. The exponent behaves as $\int_0^{\Lambda} \frac{1-\cos\omega\tau}{\pi\eta^*\omega}=\frac{1}{\pi\eta^*}\ln \Lambda\tau$, since the $\tau$ integral is dominated by long $\tau$, hence
\begin{equation} \label{e07}
 \eta^*\omega=\frac{2}{\pi}\eta\Lambda^{2-2K-1/\pi\eta^*}\int_0^{\infty}
 d\tau\frac{1-\cos\omega\tau}{\tau^{2K+1/\pi\eta^*}}=\eta\omega
\end{equation}
which is a consistent solution on a line $\frac{1}{\pi\eta}=2(1-K)$.
The third possible solution is similar to the bare one $f(\omega) = M^*\omega^2$, then the exponent behaves as
$\int_0^{\infty}\frac{1-\cos\omega\tau}{\pi M^*\omega^2}=|\tau|/2M^*$, leading to $M^*\approx M$ for intermediate or weak coupling.
The variational scheme can be shown to be related to an RG process \cite{horovitz_aharonovbohm_dissipative} from which the fixed point line Eq. (\ref{e03}) is reproduced. We note that both RG and the variational method are valid as weak coupling expansions where the coefficients in Eq. (\ref{e03}) are small, i.e. large $\eta$ and small $|1-K|$.

The above methods lead thus to three different possible behaviors for the system: \\
i) At $1-K=\frac{1}{2\pi\eta}$ the particle propagator has the friction form $(\eta |\omega|)^{-1}$, i.e. the nonlinearity of the cosine and the long range effect balance each other to produce an equivalent action with $\eta|\omega||X(\omega)|^2$. \\
ii) The case
$1-K<\frac{1}{2\pi\eta}$ flows to small $\eta$ and eventually to an $M^*\omega^2$ form, with $\langle(X_\tau-X_0)^2\rangle \sim |\tau|$, which corresponds to a delocalized phase. The effective mass $M^*$ is identified by the RG flow, as in Eq. (\ref{e04}).
Note that even in this delocalized phase, some effects of the underlying quasi-long range periodicity of the LL with the wavevector $2\pi\rho_0$ are still felt by the particle. Indeed its correlation \emph{at that periodicity} are only very slowly decaying
$\langle\cos X_\tau\cos X_0\rangle\sim \tau^{-2K}$. This indicates that the particle has a much greater chance to be found at some particular places on the chain. This can be understood qualitatively by the argument that the particle moves in the
``charge density wave'' of wavevector $2\pi\rho_0$ provided by the LL, hence the particle diffuses predominantly by tunneling between lattice sites spaced by $1/\rho_0$. On the mathematical side, this property which
is apparent in a first order calculation in $\eta$ \cite{horovitz_aharonovbohm_dissipative} is in fact known in general in the context of XY models with long-range interactions \cite{spohn_XY_longrange}.\\
iii) The case $1-K>\frac{1}{2\pi\eta}$
flows to large $\eta$ with eventually $f(\omega)\sim\omega^{2K-1}$ i.e. $S_0$ of Eq. (\ref{e02}) is a fixed point action.  From this form one could naively expect that the correlations of $\langle[X_\tau-X_0]^2\rangle$ to be convergent and thus this phase to be a localized one. The situation is in fact more subtle and we discuss this phase in more details below.

A summary of the various regimes can be found in Fig.~\ref{fig:phasediag} and the corresponding correlation functions are indicated in Table~\ref{tab:corr}.  At finite temperatures $T$ and after analytic continuation to the retarded response at real time $t$ \cite{giamarchi_book_1d} we find the replacements $|\tau|^{-2K}\rightarrow \sin \pi K \eexp{-2K\pi T t}$, $|\tau|^{2K-2}\rightarrow \sin \pi (1-K) \eexp{-(2-2K)\pi T t}$ and $\ln|\tau|\rightarrow Tt/\eta$, i.e. diffusion in real time on the dissipative line.

\begin{table*}
\caption{\label{tab:corr} Correlations of the phases in Fig. 1 at $T=0$.}
\begin{center}
\begin{tabular}{| c|c|c|c| c|}
\hline
correlation &   delocalized & dissipative & \hspace{2mm} periodic \hspace{2mm}   & \hspace{2mm} localized \hspace{2mm}\\
\hline
$\langle\cos X_\tau\rangle$ &  0 & 0 & constant  &  1 \\
$\langle\cos X_\tau\cos X_0\rangle$ & $\sim |\tau|^{-2K}$ & $\sim |\tau|^{-(2-2K)}$& constant   &  1 \\
$\langle (X_\tau-X_0)^2\rangle$  & $\sim |\tau|$ & $\sim \ln |\tau|$ &    &   0  \\
\hline
\end{tabular}
\end{center}
\end{table*}

For $K<1$ we complement the above analysis by a mean-field approach similar to the one used in the context of XY models with long-range interactions \cite{fisher_XY_longrange}. We take $h=\langle\cos X_\tau\rangle$ as an order parameter. The interaction term in (\ref{e01}) decouples as $\eta\Lambda^{2-2K}h\int_\tau\cos X_\tau\int_{\tau'}|\tau-\tau'|^{-2K}=\eta\Lambda h\frac{1}{2K-1}\int_\tau\cos X_\tau$. The self consistency relation, linear in $h$, is $1=\eta\Lambda\frac{1}{2K-1}\int_{\tau'}\langle \cos_\tau\cos_{\tau'}\rangle_0=4\eta M\Lambda\frac{1}{2K-1}$; it yields the critical line $\eta_c=\frac{2K-1}{4M\Lambda}$ above which $\langle\cos X_\tau\rangle\neq 0$. We expect the mean field result to be more reliable near $K=\half$, where the range of the interaction increases. As $K$ increases from $K=\half$ fluctuations will increase the critical value, eventually joining the transition line with the variational form $\eta_c=\frac{1}{2\pi(1-K)}$ near $K=1$. Note that mean field exponents become valid \cite{fisher_XY_longrange} when $K<3/4$, e.g. $\langle\cos X_\tau\rangle\sim\sqrt{\eta-\eta_c}$.
In Fig.~\ref{fig:phasediag} we plot the transition line as an interpolation between the mean field at $K<3/4$ and the variational form at $K>3/4$. We see that the point $K=1/2$ plays an important role, not captured by the variational or RG approaches. Below this point the interaction is so-long range that an ordered phase would exist, within the mean-field solution, for arbitrary strength of the coupling $\eta$.

In the periodic phase, instanton excitations must a priori be considered since one would have many degenerate minima of the order parameter.
Such instantons are known for the $K=1$ case \cite{korshunov87_dissipative_qm,bulgadaev_XY_instantons}.
Assuming an instanton with width $\tau_0$, the interaction term in (\ref{e01}) has the form $ \eta(\Lambda\tau_0)^{2-2K}B_K$  while the mass term is $\sim M/\tau_0$, hence the action is minimized at $\Lambda\tau_0\sim (\frac{M\Lambda}{(1-K)\eta B_K})^{\frac{1}{3-2K}}$ for $K<1$; the numerical prefactor $B_K$ is known at $K=1$, $B_1=\pi$. Note that the mass term and $K\neq 1$ set a finite scale for $\tau_0$, unlike the $K=1$ case.  The instanton action is then
\beq{12}
S_{inst}\approx M\Lambda \left(\frac{(2-2K)\eta B_K}{M\Lambda}\right)^{\frac{1}{3-2K}}\frac{3-2K}{2-2K}
\eeq
Such instantons mean that the coordinate $X_\tau$ can tunnel between neighboring minima of the ordered $\langle\cos X_\tau\rangle$. Assuming independent instantons this would imply that
 $\langle (X_\tau-X_0)^2\rangle =D|\tau|$ has a finite diffusion constant, $D\sim \eexp{-S_{inst}}$.

In particular we consider $K\rightarrow \half$ and an instanton localized at $\tau=0$. The dominant contribution for the instanton center at $|\tau|<\tau_0$ comes from $|\tau'|>\tau_0$ that involves $|X_\tau|\gg |X_{\tau'}|$ and $\int_{|\tau'|>\tau_0}|\tau'|^{-2K}\sim \frac{1}{2K-1}$ which diverges at $K\rightarrow \half$, hence
\beq{13}
S(K\rightarrow \half)
&&=\half M\int_{\tau}(\dot X)^2_{\tau}+\nonumber\\&&\frac{\eta\Lambda(\Lambda\tau_0)^{1-2K}}{\pi(2K-1)}\int_{\tau}
(1-\cos X_\tau)+S'
\eeq
$S'$ comes from the instanton tails where $X_\tau,X_{\tau'}$ are small (up to $2\pi$) and comparable.
This action is similar to the well known sine-Gordon system, identifying $B_K\sim (2K-1)^{-1}$ whose instanton (or soliton) solution has a width
$\tau_0\sim (2K-1)^{\half}$ and action $S_{inst}\sim (2K-1)^{-\half}$.
Assuming independent instantons the diffusion constant would diverges at $K=\half$, i.e. $\ln D\sim (2K-1)^{-\half}$. We propose that the whole range of the periodic phase in Fig.~\ref{fig:phasediag} has instanton solutions with a finite action, with an explicit solution provided by the sine-Gordon system at $K\rightarrow 1/2$.  However, given the long range form of the interaction within the tail term $S'$, to ascertain the correct behavior of $\langle (X_\tau-X_0)^2\rangle$ at large time requires
further study of how these instantons interact, which is left for the future.

We consider next the system at $K<1/2$. This case has been studied in the context of discrete XY models \cite{campa_XY_longrange,tamarit_XY_longrange,vollmayr-lee_kac_longrange} and was shown to have a phase transition in the limit that the coupling vanishes as a power of the system size, which in our case is $\beta=1/T$, i.e. there is a critical value for $\eta(\beta)^{1-2K}$. Hence at $T=0$ the system is fully ordered and $\langle\cos X_\tau\rangle=1$. Furthermore, instanton excitations would involve the effective coupling $\eta(\beta)^{1-2K}$, hence will have diverging action. Extending the mean-field analysis to $K<\half$ yields a critical temperature $T_x$ where $(1-2K)(T_x/\Lambda)^{1-2K}=2M\Lambda\eta$;  fluctuations would render $T_x$ into a sharp crossover temperature.

Let us conclude this part by noting that the hypothesis made at the beginning to neglect $X_\tau-X_0$ compared to $\tau$ is indeed justified in all the phases. Furthermore, note that although the results of the present paper are derived in the large-$N$ limit, we of course expect them to extend to a finite number of component as well. E.g. for the Coulomb box case deviations due to finite $N$ appear at exponentially small temperatures \cite{zarand_instantons_quantum_dots}.

Finally we discuss possible realizations of our model with various collective coordinates that are potential candidates for experimental studies. \\
(i) A first example that yields our action (\ref{e01}) is a fermion Coulomb box \cite{schon_josephson_review}. Following the Ambegaokar-Eckern-Sch\"on mapping
\cite{ambegaokar_josephson_dissipation_short} one introduces a phase $X_\tau$ such that $\dot X_\tau$ measures the charge
in the box while the charging energy corresponds to $1/M$. The kernel in (\ref{e01}) is then $\sum_{\alpha}G_{\alpha,i}(\tau-\tau')\sum_kG_{k,i}(\tau'-\tau)$ where $i$ is the channel index, $\alpha,k$ are internal quantum
numbers of the dot and LL, respectively, and the Green's functions are for either free fermions on the dot, $\sim 1/(\tau-\tau')$ or
for fermions in the LL (with Luttinger parameter $K_f$)  $\sim |\tau-\tau'|^{-\half(K_f+1/K_f)}$. Hence an effective action of the form (\ref{e01}) with $2K=1+\half(K_f+1/K_f)$,  realizing only $K>1$ cases.\\
(ii) A variation of realization (i) is a system of LL that terminate in a Coulomb box, i.e. a region where all LL have long range Coulomb interaction with an effective capacitance, as illustrated in Fig. 1a.  In this case boundary Green's function \cite{giamarchi_book_1d} is needed $G_{x=0,i}(\tau-\tau')\sim|\tau-\tau'|^{-1/K_f}$, hence $K=\frac{1}{K_f}$  and the interesting regime  of Fig. 1 with $K<1$ is realizable with attractive interactions $K_f>1$. In case that Coulomb box region is a normal metal we obtain $2K=1+\frac{1}{K_f}$ .\\
(iii) A 3rd realization  (Fig. 1b) corresponds to a BEC with a phase $\theta_\tau$ that weakly couples to bosonic LL's with boson operators $\Psi_n(\tau)$ as $g\eexp{i\theta_\tau}\Psi_n(\tau)+h.c.$. The average involves now the boson's Greens function $\sim |\tau-\tau'|^{1/2K_b}$, $K_b\rightarrow \infty$ for noninteracting bosons and $K_b$ decreases to $1$ for on site repulsion $U\rightarrow\infty$. Hence (\ref{e01}) is realized with $K=1/4K_b$ and the localized regime (Fig. 1) with $K<\half$ is realizable.\\
 (iv) In analogy with BEC a superconducting grain can Josephson couple to superconducting one-dimensional wires (Fig. 1b). For attractive short range inteactions $2K=\frac{1}{K_\rho}$ and $K<1$ can be realized by fermions (spinfull in this example) with long range repulsive or attractive interactions allowing for the interesting regime $K<1$. This case could potentially be realized with the new superconducting LaAlO$_3$/SrTiO$_3$ nanostructures \cite{stornaiuolo}\\
 (v) Finally, the mobile impurity case may be realized by an impurity confined in between LL chains forming e.g. a hexagon. In this case the interesting $K<1$ regime is realized by repulsive fermion interactions.

In conclusion, we have studied the physics of LL environments that couple to a collective coordinate such as an impurity position, charge of a Coulomb box, a phase of a BEC or that of a superconducting grain. We have shown that the coupling to the bath leads to various phases for the collective coordinate ranging from delocalized, dissipative, periodic and localized.
Our results are summarized in Fig.~\ref{fig:phasediag} and Table~\ref{tab:corr}, showing the distinctions among the various phases.
We believe that the large set of realizations for the collective coordinate and the various phase transitions will stimulate further research.

Acknowledgments: We thank E. Demler, E. Berg, E. Dalla Torre, B. Halperin, A. Kamenev, C. Mora  and A. Tsvelik for stimulating discussions. This work was supported in part by the Swiss NSF under MaNEP and Division II. BH acknowledges kind hospitality and support from CMT at Harvard, from DPMC-MaNEP at University of Geneva, from the Institut Henri Poincar{\'e} and from LPT at Ecole Normale Sup{\'e}rieure.
TG is grateful to  the Institut Henri Poincar{\'e}, the Harvard Physics
department and the MIT-Harvard Center for Ultracold Atoms for
support and hospitality.
PLD thanks ANR grant 09-BLAN-0097-01/2.



%

\end{document}